\documentclass[sigconf, review=false,  numbers]{acmart}
\usepackage{float}
\usepackage{graphicx}
\usepackage{tabularx}
\usepackage{pdflscape}
\usepackage{placeins}
\usepackage{lineno}
\setcopyright{none}
\acmYear{}
\acmMonth{}
\acmConference{}{}{}
\acmISBN{}
\acmDOI{}
\setcopyright{none}
\AtBeginDocument{%
  }
\usepackage{natbib}
\begin{document}
\nolinenumbers

\title{SightGlow: A Web Extension to Enhance Color Perception and Interaction for Vision Deficiency}

\author{Sansrit Paudel}
\email{sansrit@uri.edu}
\orcid{0009-0004-4309-2870}
\authornotemark[1]
\affiliation{%
  \institution{University of Rhode Island}
  \city{South Kingstown}
  \state{Rhode Island}
  \country{USA}
}

\begin{abstract}
SightGlow is a web extension tailored to improve color perception accuracy for individuals with red-green color blindness. The research was focused on evaluating whether personalized color adjustment and selective zoom enhance user interaction and satisfaction for individuals with low vision and color vision impairment. The system was developed as an iterative process by conducting a pilot
user survey. Existing web extensions were limited in addressing challenges faced by low vision and color blindness, hence this application provides additional features including selective zoom and
color controls, which make this unique. Most participants responded that the application’s flexibility to adjust the color balance for
any images or video graphic content enhanced their user experience hence resulting effectiveness of the system.
  
\end{abstract}


\maketitle

\section{Introduction}
Color Blindness is a color vision deficiency that affects individuals ability to see and differentiate between certain colors \cite{10.1145/3681756.3697945}. Color vision is determined by three different color components: hue, saturation, and brightness.The fundamental difference between color blindness and normal people is a color that looks different to most people whereas same to color blind people. In other words, having defective color vision means the ability to distinguish color, saturation and brightness decreases. \cite{article} Statically color blindness is one of the global problem which exits in 1 out 12 men affecting (8\%) and 1 in 200 women around (4.5\%), world wide as a result more than 350 millions color blind people in the world  \cite{colorblind_population_counter}. There exists 4 variety of color blindness, and most prevalent form of color blindness are red-green color blindness which I aim to explore. Despite the challenges it presents, individuals
with color blindness develop strategies to compensate this limitation in their daily activities. However, the condition can restrict access to certain occupations and activities that rely heavily on color discrimination \cite{byrne2010things}. Sight-Glow was therefore introduced to solve two distinct problem, 1st to contribute on making social media more engaging and fun thereby  enhancing color perception on videos and images in real time. Secondly to provide a selective zoom functionality to work with text heavy document. The functionality not only helps them reading documents it further makes social media engaging to let them read comments with more flexible way. 
\section{Related Work}
Multiple papers have presented a framework for correcting color documents so that they are accessible to color-blind viewers. Upon literature review, several mathematical models have been proposed that can be implemented to achieve the goal. The algorithm introduced contains a way to map colors using the World Wide Web Consortium evaluation criteria so that detail is pre-served for color-blind viewers, especially dichromats.\\ \\
\citeauthor{10.1145/1240624.1240855} (2007) have developed adaptation algorithm for improving the accessibility to color images for color vision deficiency computer users. They have worked on algorithm that transfers the chromatic information of the defective cone across the two functioning cones. This technique to shift the color information from weak color receptor to working channel provided better color perception \cite{10.1145/1240624.1240855}.\\\\
\citeauthor{article} (2019) designed video streaming application especially tailored to color-blind with optimized color palettes including labeled filters and contrast adjustment to enhance visual clarity \cite{article}.\\
\citeauthor{10.1145/2632048.2632091} have developed a system named Chroma especially designed for augmented reality where individual with color blind can visualize the virtual environment with improved color perception. They have implemented Daltonization algorithm, a color correction technique, which can simulate how color blind individual perceive colors and then adjusting those colors to make color more distinguishable. 
\section{Research Statement}
With the primary goal on enhancing the web experience for people with low vision and color blind the application was purposed and built. The fundamental aim of this research was to find an answer on the following sets of questions.
\begin{itemize}
    \item Does selective zoom and personalized color enhances readability and provide higher visual perception among individuals with low vision and color blindness?
    \item Can SightGlow enhance user experience with its personalized solutions for video-graphic content? 
\end{itemize}
\section{System Design and Methodology}
\subsection{User Interface}
The user interface was initially prototyped taking account the user's need with red-green color-blind. Individuals with red-green color blind can better perceive either blue or orange color \cite{Color_Blue_yellow} palette. Hence, particular colors were explicitly selected for creating front end design on my system. 
\begin{figure}[H]
    \centering
    \includegraphics[width=0.5\linewidth]{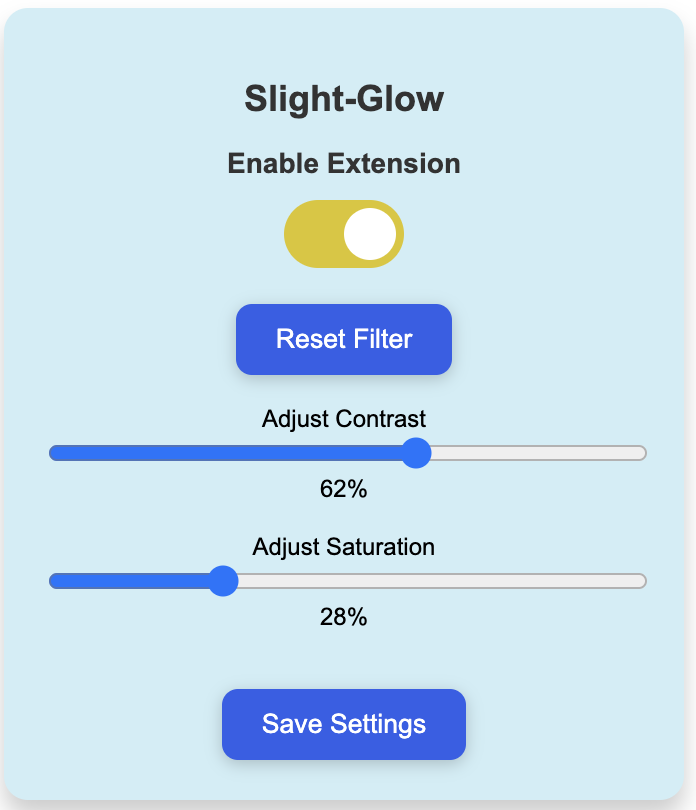}
    \caption{User Interface SightGlow.}
    \Description{A screenshot of the SightGlow user interface showing a color blind friendly design using blue and orange colors.}
    \label{fig:User Interface SightGlow}
\end{figure}

As shown in Figure ~\ref{fig:User Interface SightGlow}, the color in the user interface was optimized for red-green color-blind. The larger toggle button and slider were designed to let the user control the functionality with fewer motor actions.

\subsection{Color Filters}
Color perception is achieved through cone cells in the retina. Humans with normal vision have 3 types of cells, sensitive to different light wavelengths. L cones capture Long-wavelength (Red), M cones capture medium-wavelength (green), and S cones capture Short-wavelength (Blue) \cite{daltonlens}. 
\begin{figure}[H]
    \centering
    \includegraphics[width=0.5\linewidth]{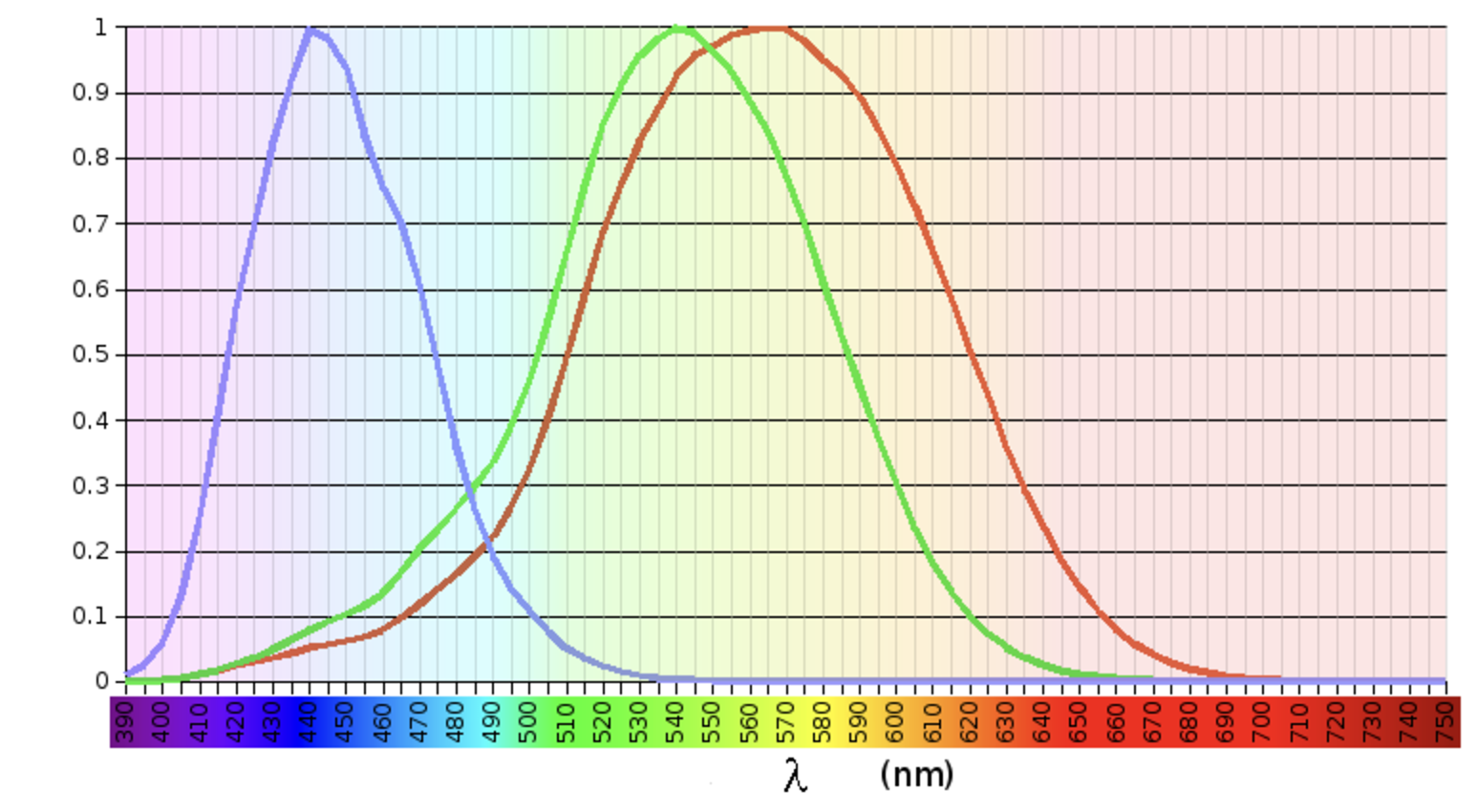}
    \caption{Normalized responsivity spectra of human cone cells, S, M, and L types.}
    \Description{A screenshot of the SightGlow user interface showing a color blind friendly design using blue and orange colors.}
    \label{fig:User Normalized responsivity spectra of human cone cells, S, M, and L types}
\end{figure}
The color correction is acquired by converting RBG color space to LMS to simulate particular color blindness and applying daltonizing algorithm to providing most alike color to the targeted color. For example individual with red color blindness doesn't perceive color red however daltonizing algorithm reproduces colors converting color red to pink scheme which that helps to color-blind to distinguish two contrasting color tone. Hence, the total process includes transforming the RGB image into LMS space reproducing the simulation, applying color matrix for adjustment and retracing LMS space into RGB.\\\\

In-order to apply the most optimal color filter a specific reference filter matrix applied which was computed through  Daltonization algorithm. 
\begin{itemize}
    \item The matrix  (5x5) would be computed for output colors ($R'$, $G'$, $B'$, $A'$).
    \item The columns (5x1) matrix represents how the input color channels ($R$, $G$, $B$, $A$) and a constant bias term contribute to the output.
\end{itemize}

\subsection{Matrix Representation}
The original color space were transformed with new matrix multiplication to acquire desired filter overlay. Here, [4x5] matrix is introduced which daltonizes the original color. 
\[
\begin{bmatrix}
R' \\
G' \\
B' \\
A'
\end{bmatrix}
=
\begin{bmatrix}
r_1 & r_2 & r_3 & r_4 & r_5 \\
g_1 & g_2 & g_3 & g_4 & g_5 \\
b_1 & b_2 & b_3 & b_4 & b_5 \\
a_1 & a_2 & a_3 & a_4 & a_5
\end{bmatrix}
\cdot
\begin{bmatrix}
R \\
G \\
B \\
A \\
1
\end{bmatrix}
\]

Where:
\begin{itemize}
    \item $R'$, $G'$, $B'$, $A'$: Represents the resultant color matrix, each representing the color channels (Red, Green, Blue, Alpha) respectively.
    \item $R$, $G$, $B$, $A$: Represents the input color channels (Red, Green, Blue, Alpha) respectively.
\end{itemize}
The multiplication of [4x5] matrix with input input color channel, [5x1] results in creating the daltonized graphics. Each row in the matrix results to how a specific output channel is calculated from the input channels\cite{desmic2024}. Following section provides description how matrix multipication is carried out and results are directed. 
\[
R' = (0.5 \cdot R) + (0.2 \cdot G) + (0.3 \cdot B) + (0 \cdot A) + 0
\]
\[
G' = (0.0 \cdot R) + (1.0 \cdot G) + (0 \cdot B) + (0 \cdot A) + 0
\]
\[
B' = (0.2 \cdot R) + (0.3 \cdot G) + (0.5 \cdot B) + (0 \cdot A) + 0
\]
\[
A' = (0.0 \cdot R) + (0.0 \cdot G) + (0.0 \cdot B) + (1 \cdot A) + 0
\]
Daltonizing matrix that is currently used to our system  for red-green color blind: \\
\[ 
\centering
\begin{bmatrix}
0.5 & 0.2 & 0.3 & 0 & 0 \\
0.0 & 1.0 & 0 & 0 & 0 \\
0.2 & 0.3 & 0.5 & 0 & 0 \\
0.2 & 0 & 0.0 & 1 & 0
\end{bmatrix}
\]
Here, the resultant output is a weighted combination of original RGB color values. The opacity, or apha channel is kept unchanged, as it does not alters any color perception.
\section{System Flow Diagram}
Following system flow diagram would provide holistic overview of our system. 
\begin{figure}[H]
    \centering
    \includegraphics[width=0.5\linewidth]{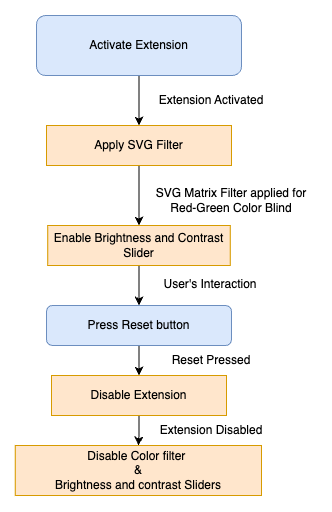}
    \caption{System flow diagram.}
    \Description{Descriptive system flow diagram.}
    \label{fig:User Normalized responsivity spectra of human cone cells, S, M, and L types}
\end{figure}
The system when enabled adopts a daltonizing filter matrix which would transform the existing screen body with new color channel. This activation enables the system to further allowing users to use selective display functionality, which provides functionality to select any text content to visualize text on a  popup card with increased font size. 
\begin{figure}[H]
    \centering
    \includegraphics[width=0.5\linewidth]{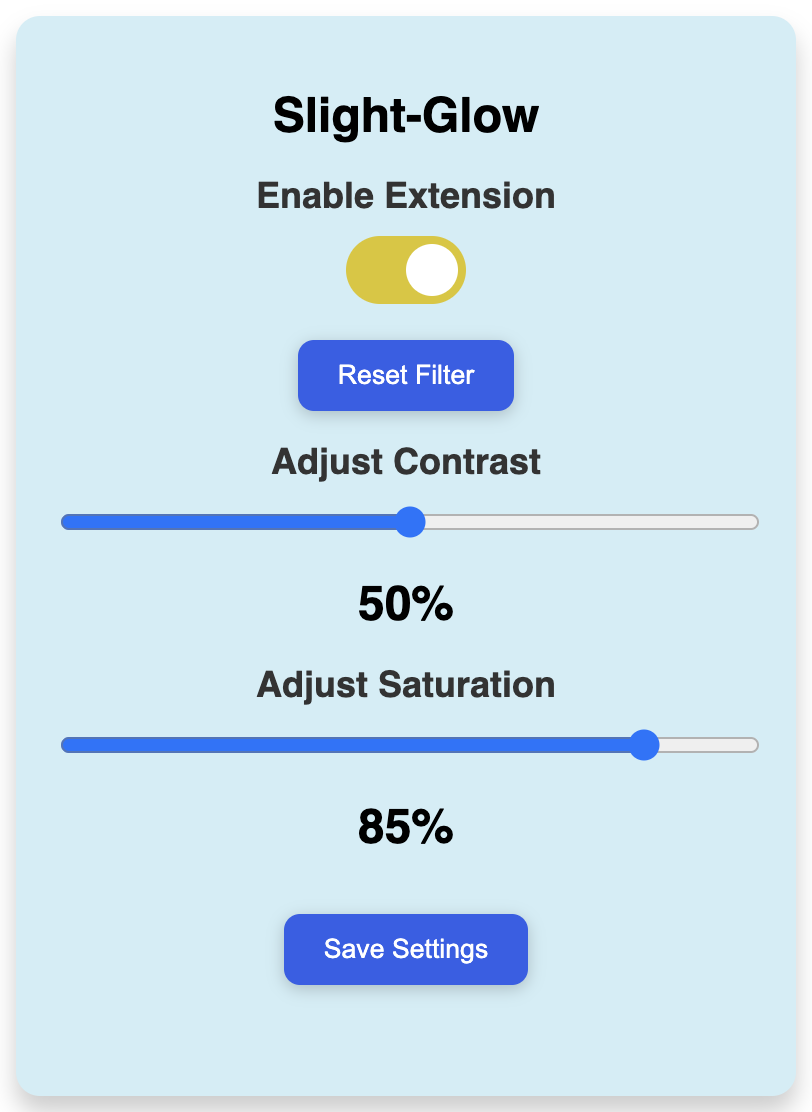}
    \caption{Redesigned interface.}
    \Description{Descriptive system flow diagram.}
    \label{fig:Redesigned system after use's evaluations}
\end{figure}
The system provides multiple functionalities to implement such as daltonizing color matrix ,selective zoom and further to adopt personalized experience. 
\section{Features}
This application provides 3 major features which includes:
\begin{enumerate}
    \item SVG color filter tailored to color blind.
    \item Selective zoom.
    \item Personalized color adjustment.
\end{enumerate}
\subsection{Color Filter}
People with no vision deficit perceives true color whereas color-blind perceives in more divergent way.
\begin{figure}[H]
    \centering
    \includegraphics[width=0.5\linewidth]{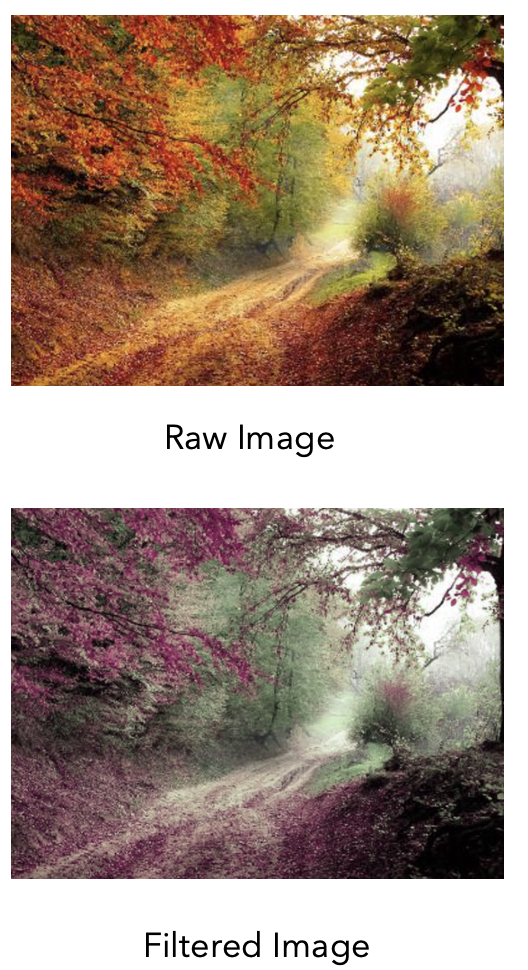}
    \caption{Top image represents raw colored image whereas button represents daltonized.}
    \Description{Descriptive system flow diagram.}
    \label{fig:User Normalized responsivity spectra of human cone cells, S, M, and L types}
\end{figure}
\begin{figure}[H]
    \centering
    \includegraphics[width=0.5\linewidth]{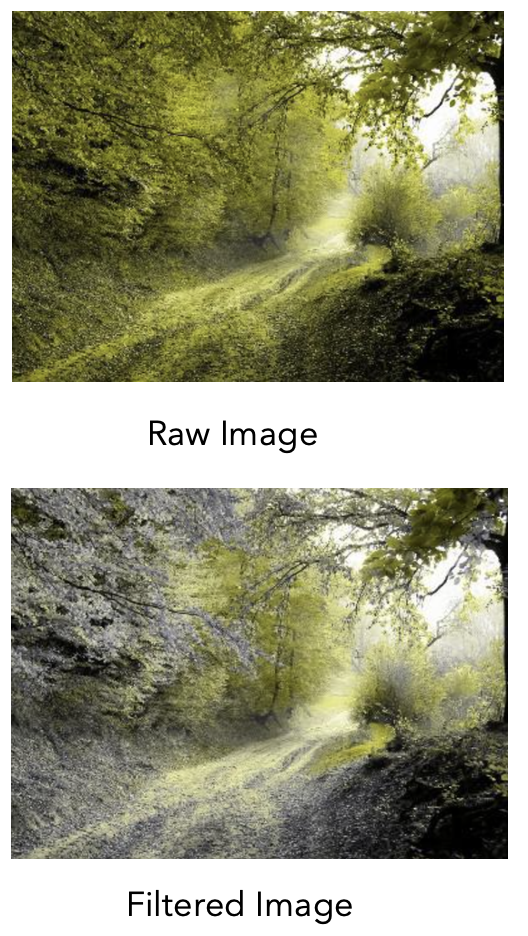}
    \caption{Top image represents simulated image representing red-green color blind, whereas button represents daltonized image perceived by color blind.}
    \Description{Top image represents raw colored image whereas button represents daltonized}
    \label{fig:How color blind perceives color}
\end{figure}
Creating assistive technology for color blind, doesn't necessarily mean that we can present them true color, it's the nerve function which can't be altered by software solution however we can help them better differentiate between two distinct color layout which they perceive alike.
\subsection{Selective Zoom}
\begin{figure}[H]
    
    \includegraphics[width=0.9\linewidth]{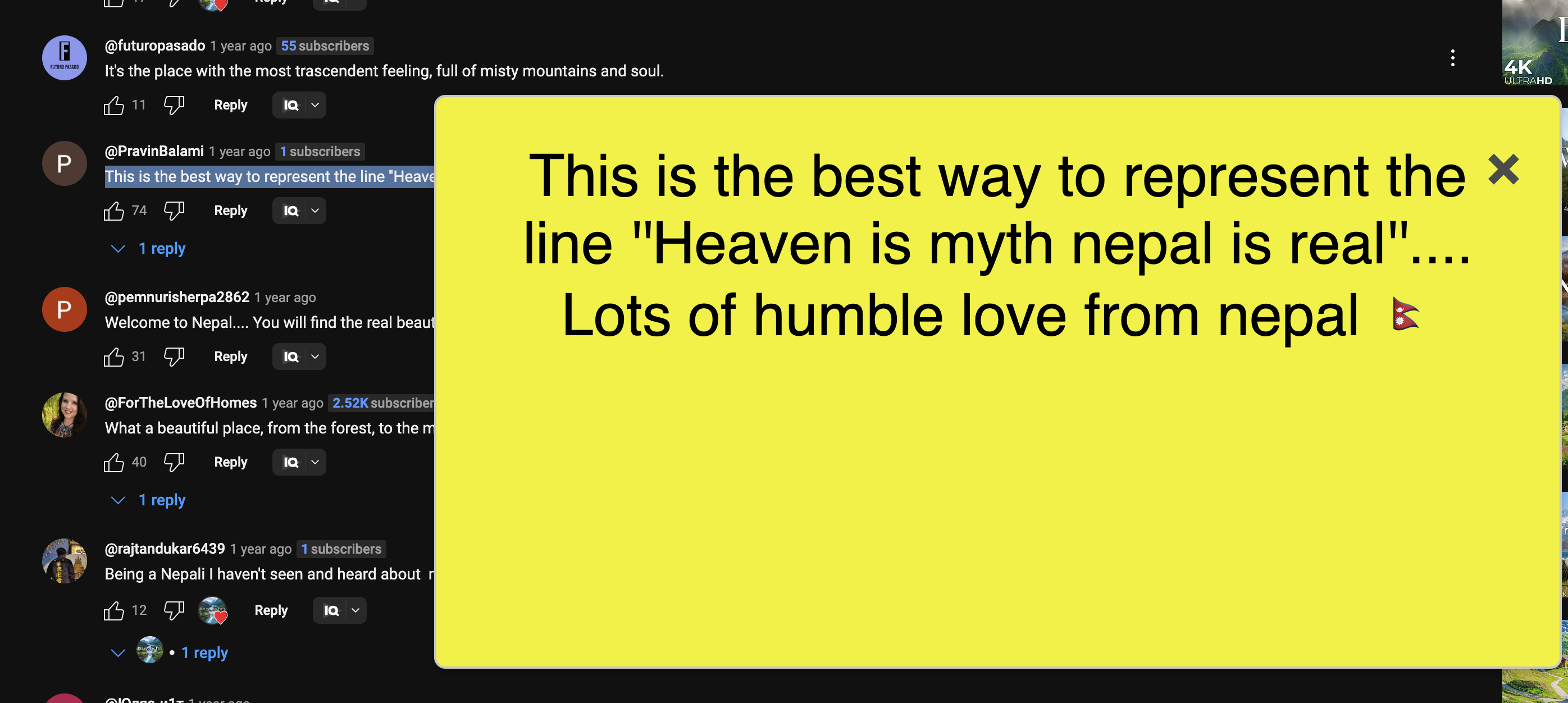}
    \caption{Selective zoom functionality}
    \Description{Selective zoom functionality}
    \label{fig:How color blind perceives color}
\end{figure}
Selective zoom is an application featured where users can select any text into their web interface which returns them with a popup card with enlarged text content, for enhancing readability. Color yellow was particularly selected considering the color perception accuracy amoung color-blind. 
\subsection{Optimizing Color }
\begin{figure}[H]
    \centering
    \includegraphics[width=0.9\linewidth]{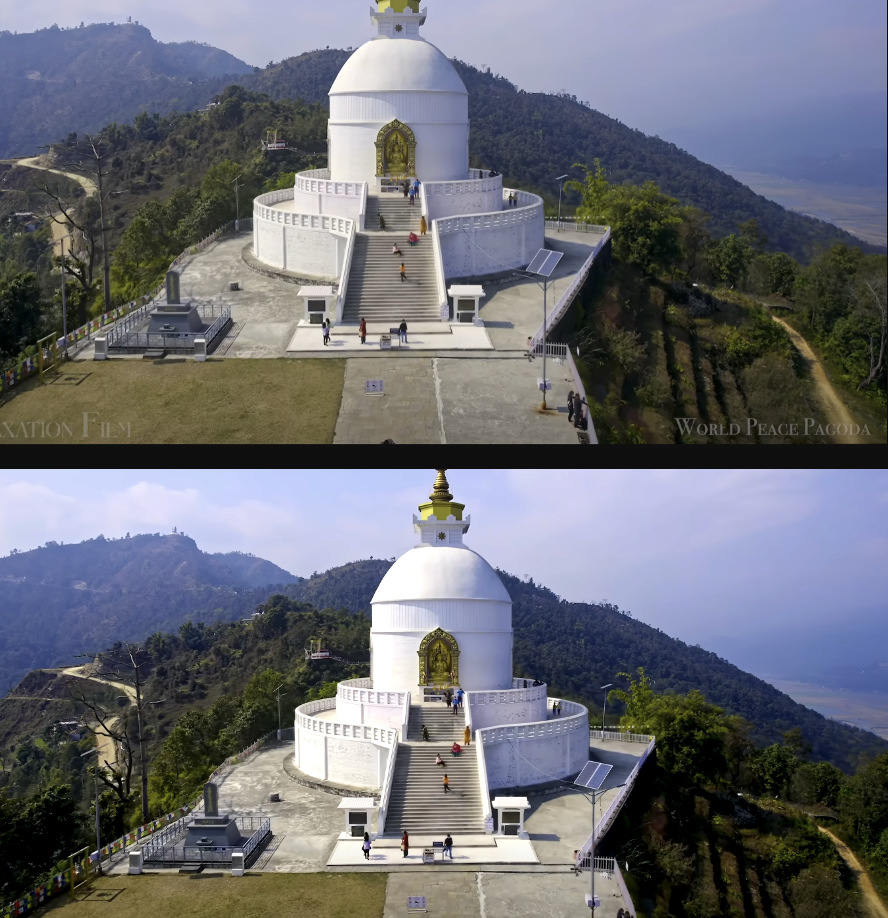}
    \caption{Top image represents images from  video clip, button represents personalized color for better visibility }
    \Description{Dynamic control for color adjustment was implemented.}
    \label{fig:Dynamic color control for enhanced visibility}
\end{figure}
Slider were designed and it's functionality provides dynamic color control. Contrast and saturation can be altered for images and videos to adjust user's need.This provided real-time adjustment, enhancing color perception and accuracy on images and videos.
\section{Usability Evaluation}
In order to evaluate the extension, usability evaluation was conducted.The objective of user study was to early identify potential design flaws, user experience and receive inputs from user's perceptive. Firstly, nine individuals with normal vision participated on my case study. Seven multiple choice questions on Likert scale were presented with one feedback question, for the initial evaluation.These includes
\begin{itemize}
    \item Q1.The extension was easy to use and understand.
    \item Q2.The selective zoom feature enhanced readability.
    \item Q3.The popup card with large fonts was visually comfortable-
able.
    \item Adjusting color contrast and saturation enhanced my
experience with videos.
    \item Q5.The extension helped with adjusting the color accuracy of my
choice.
    \item Q6.The graphical interface of the system was visually appealing-
ing.
    \item Q7.Overall, I am satisfied with the developed system.
\end{itemize}

\subsection{Findings}
Upon user's evaluation from pilot study, I have received a mixed feedback where most response were positive. The useful finding from participants were, they highlighted the potential flaws in interface design, being font size much smaller. The system was tailored to low vision hence the feedback was very constructive and meaningful. Apart from that some participant expressed their thought to have the system to work on PDF documents. 
\subsection{Iterative Design}
Considering user's feedback and guidelines some changes were made into the sytem. The interface was redesigned with larger fonts and layouts, which addressed the user's needs, which is represented on figure 4.

Does selective zoom and personalized color enhances read-
ability and provide higher visual perception among individ-
uals with low vision and color blindness?
• Can SightGlow enhance user experience with its personal-
ized solutions for video-graphic content?
\section{Results and Conclusion}
The system was developed as an iterative process in-cooperating user's feedback to enhance its functionalities. A systematic pilot study was conducted with nine participants to evaluate user's experience and effectiveness of the system. The response were positive with most participants finding the interface visually appealing and features significantly enhanced their reading abilities. Selective zoom was found to be intriguing to participant and the majority of participants provided positive response on it's effectiveness. While a few participants were neutral however overall responses were very encouraging. Few participants were not convinced that application enhanced the readability while they were positive on color controls the system provides. Three participants provided constructive feedback on testing the system. Looking at a border impact this application can benefit senior citizen with short slightness and low vision, who love to engage in social media and spend their solitary life. The system test with people with colour blindness is yet to be done. Future works includes testing the effectiveness of the application with colour-vision users and evaluating the device's effectiveness. 
\begin{acks}
This mini research project was completed as a part of my coursework "CS592-Designing Interactive System". I extend my sincere gratitude to Professor Shaun Wallace, for his critical guidance  and valuable insights throughout the semester.\\\\
\end{acks}
\appendix
\section{Appendix}
List of abbreviations and acronyms:
\begin{table}[H]
\centering
\begin{tabular}{|c|l|p{4cm}|}
\hline
\textbf{S.N} & \textbf{Abbreviation} & \textbf{Description} \\ \hline
1            & RGB                   & Red, Green, Blue \\ \hline
2            & LMS                  & Large, Medium, Small\\ \hline
\end{tabular}
\caption{List of Abbreviations and Descriptions}
\label{tab:abbreviations}
\end{table}

\bibliographystyle{ACM-Reference-Format}
\bibliography{sightGlow}

\end{document}